\pdfoutput=1
\documentclass[12pt]{iopart}

\usepackage{iopams}  

\usepackage{graphicx,xr}

\newcommand{\nvec}[1]{{\mathbf #1}}
\begin{document}

\title[Geometry dipole-dipole]{Geometrical simplification of the dipole-dipole interaction formula}

\author{Ladislav Kocbach and Suhail Lubbad}

\address{Dept. of Physics and Technology, University of Bergen, Norway }
\ead{ladislav.kocbach@ift.uib.no   suhail.lubbad@gmail.com}
\begin{abstract}
Many students meet quite early this dipole-dipole potential energy
when they are taught 
electrostatics or magnetostatics, and it is also a very popular formula, 
featured in the encyclopedias. We show that by a simple rewriting of the formula 
it becomes apparent that for example, by reorienting the two dipoles, their
attraction can become exactly twice as large. The physical facts are naturally 
known, but the presented transformation seems to underline the geometrical features in 
a rather unexpected way. The consequence of the discussed features is the so 
called magic angle which appears in many applications. The present 
discussion also contributes to an easier introduction of this feature.
We also discuss a possibility for designing educational toys and try to suggest
why this formula has not been written down frequently before this work. Similar 
transformation is possible for the field of a single dipole, there it seems to be 
observed earlier, but also in this case we could not find any 
published detailed discussion. \\
\end{abstract}
\maketitle
\parindent 0cm
\parskip 0.25cm
%
%%%%%%%%%%%%%%%%%%%%%%%%%%%%%%%%%%%%%%%%%%%%%%%%%%%%%%%%%%%%%%%%
%
\section{Introduction  \label{sect_intro}}
%
%%%%%%%%%%%%%%%%%%%%%%%%%%%%%%%%%%%%%%%%%%%%%%%%%%%%%%%%%%%%%%%%
%
Most people who had a chance to play with magnets are aware of the 
fact that the magnets attract each other when held in two different orientations, 
but very few will be able to establish that one of the forces of attraction
is exactly twice as large as the one in the weaker configuration. 
This simple fact is hidden in the well known formula for 
the dipole-dipole interaction.
Many students meet quite early this dipole-dipole potential energy
when they are taught 
electrostatics or magnetostatics, and it is also a very popular formula, 
featured in the encyclopedias (on the web as well as on the paper). 
In our research this formula is also used 
for certain types of molecular interactions. The dipole-dipole formula is
%
%%%%%%%%%%%%%%%%%%%%%%%%%%%%%%%%%%%%%%%
%
\begin{equation}
 U(\nvec{r}_{12}) = 
%  \frac{1}{8 \pi} 
C
\frac{1}{r_{12}^{\ \ 3}}  \left[ \nvec{m}_1
   \cdot \nvec{m}_2 - \frac{3 \left( \nvec{m}_1 \cdot \nvec{r}_{12}
   \right) \left( \nvec{m}_2 \cdot \nvec{r}_{12} \right)}{r_{12}^{\ \ 2}}
   \right]
   \label{standard_dipole_dipole}
\end{equation}   
%
%%%%%%%%%%%%%%%%%%%%%%%%%%%%%%%%%%%%%%%%
%
where $\nvec{r}_{12}$ is the vector connecting the two dipoles. The constant $ C $
is very different for magnetic and electric dipoles and we shall not specify it
at all in the present discussion. Though we discuss mainly the magnetic dipoles
because they are easily represented by small magnets, the geometrical features 
are exactly the same for the electric dipoles.

The formula looks rather uninviting, but obviously
it contains the truth about the interaction, and when needed it does its job
very well. 
Sometimes it appears as especially 
ugly species which might even
contain  different inverse powers, as $r^{-3}$ and $r^{-5}$. People
who are familiar with the model of interatomic interaction, 
the Lenard-Jones potential, 
which also features the 
difference of two powers, might then wonder if there is something
untold about minimum or maximum
Naturally, one should  remember
that in the dipole-dipole case the two powers are only an artefact of 
a sort of economic notation. The idea of the discussed transformation
probably occured to many people,  
but in our case a happy coincidence
was the presence of the two small magnets used for posting the 
anouncements on the news board, which gave the idea a practical test,
as descibed below.

The formula describes the potential energy, the forces on the bodies carrying the
dipoles and the torques on the dipoles will depend on the physical arrangements
and there are too many possibilities. In the whole paper we simply consider
only the situation that the directions of the two dipoles are kept fixed,
which means that the forces on the two bodies are given only by the gradient of the 
radial part, which is indeed very simple. The torques on the dipoles and their effects may be 
spectacular and fascinating, but they are not a part of this discussion. 
%%%%%%%%%%%%%%%%%%%%%%%%%%%%%%%%%%%%%%%%%%%%%%%%%%%%%%%%%%%%%%%%
%
%%%%%%%%%%%%%%%%%%%%%%%%%%%%%%%%%%%%%%....Figure 1
%
\begin{figure}[htb]
\center{
\includegraphics[height=8.5cm]{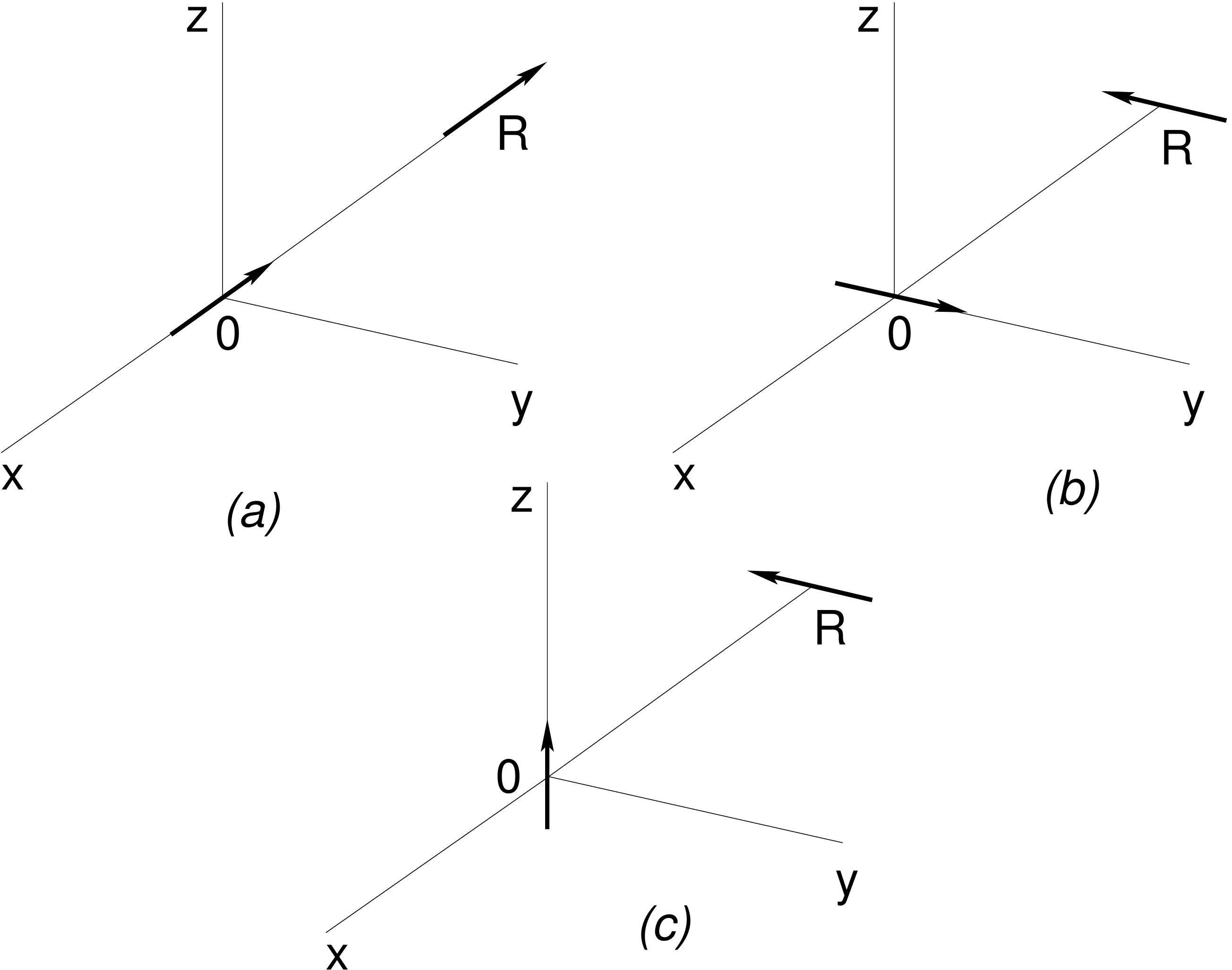}
}
 \caption{
 The magnets attracting each other when 
 positive maximum scalar product (a) due to the $-2$ term, 
 and when the scalar product is negative (b) due to the first term.
 There is a zero potential and thus zero force (c) when the
 dipols are forced to remain perpendicular.
 The dipol directions shown by 
 arrows.
}
 %%%
\label{lab.fig.general}
\end{figure}

This short note is a result of our observations, and we really think that
the suggested new form is suitable for presentation to even younger
students. Besides, we present a toy to illustrate the formula. Further, an
analogue of the trick shown here can be used to draw field lines 
of a dipole. Moreover, we think that this much more intuitive formula 
and the physical features which it so clearly displays, will be of use 
in explanations and discussions in the interdisciplinary field 
of molecular dynamics applications - and sometimes perhaps even in chemistry.

The standard form of the dipole-dipole interaction 
energy formula (repeated from eq. \ref{standard_dipole_dipole}) is 
$$
U(\nvec{r}_{12}) = 
% \frac{1}{8 \pi} 
C
\frac{1}{r_{12}^{\ \ 3}}  \left[ \nvec{m}_1
   \cdot \nvec{m}_2 - \frac{3 \left( \nvec{m}_1 \cdot \nvec{r}_{12}
   \right) \left( \nvec{m}_2 \cdot \nvec{r}_{12} \right)}{r_{12}^{\ \ 2}}
   \right] 
$$
where $\nvec{r}_{12}$ is the vector connecting the 2 dipoles. If
%
%%%%%%%%%%%%%%%%%%%%%%%%%%%%%%%%%%%%%%%%
%
\begin{equation}
\nvec{r}_{12} = r_{12}  \nvec{e}_{12} 
   \label{define_unitvector}
\end{equation}
%
%%%%%%%%%%%%%%%%%%%%%%%%%%%%%%%%%%%%%%%%
%   
we can decompose both of the dipoles into two components and write
%
%%%%%%%%%%%%%%%%%%%%%%%%%%%%%%%%%%%%%%%%
%
\begin{equation}
 m_1 = m_{1 \|}  \nvec{e}_{12} + \nvec{m}_{1 \perp} 
   = 
   \nvec{m}_{1\|} +  \nvec{m}_{1 \perp}
   \label{decompose-parallel}
\end{equation}
%
%%%%%%%%%%%%%%%%%%%%%%%%%%%%%%%%%%%%%%%%
%   
With this notation the interaction energy of equation \ref{standard_dipole_dipole}
can be written as
%
%%%%%%%%%%%%%%%%%%%%%%%%%%%%%%%%%%%%%%%%
%
\begin{equation}
U =            %   \frac{1}{8 \pi} 
C
\left( \frac{1}{r_{12}} \right)^3 
\left[ 
   \nvec{m}_{1 \perp} \cdot \nvec{m}_{2 \perp} 
   - 2   \nvec{m}_{1\|}    \cdot \nvec{m}_{2\|}
   \right]    
   \label{new_dipole_formula}
\end{equation}
%
%%%%%%%%%%%%%%%%%%%%%%%%%%%%%%%%%%%%%%%%
%   
Thus we see that the attraction can happen if the 
$   \nvec{m}_{1 \perp} \cdot \nvec{m}_{2 \perp}       $ 
is negative, i.e. pointing against each other, and then
only with factor one; or, as is the more well known fact, if the two moments
are parallel in the same direction, i.e. the factor 
$   - 2   \nvec{m}_{1\|}    \cdot \nvec{m}_{2\|}  $  
dominates.

It is naturally true that 
the standard formula does not
 need any additional definition of the perpendicular components and 
 as before should probably remain the standard notation when
 no geometrical analysis is involved.
%
%
%
%
%
%
%
%%%%%%%%%%%%%%%%%%%%%%%%%%%%%%%%%%%%%%%%%%%%%%%%%%%%%%%%%%%%%%%%
%
%
\section{Experiments with small magnets  \label{sect_Experiments_magnets}}  
%
%
%%%%%%%%%%%%%%%%%%%%%%%%%%%%%%%%%%%%%%%%%%%%%%%%%%%%%%%%%%%%%%%%
%
%
The pictures below show the two small magnets which we have by chance 
found on a magnetic board. The magnets were made in an unexpected way, 
having the south-north axis not along the length, as expected, but in 
the direction shown in the figures.

This makes it very attractive to explore the geometrical properties
of the interaction. The attraction is too strong to be able to appreciate
the difference between the "two terms", but 
the repulsion can nicely be explored when the magnets are placed on paper and 
lightly manipulated by the fingers.

\begin{figure}[htb]
\center{
\includegraphics[height=7.0cm]{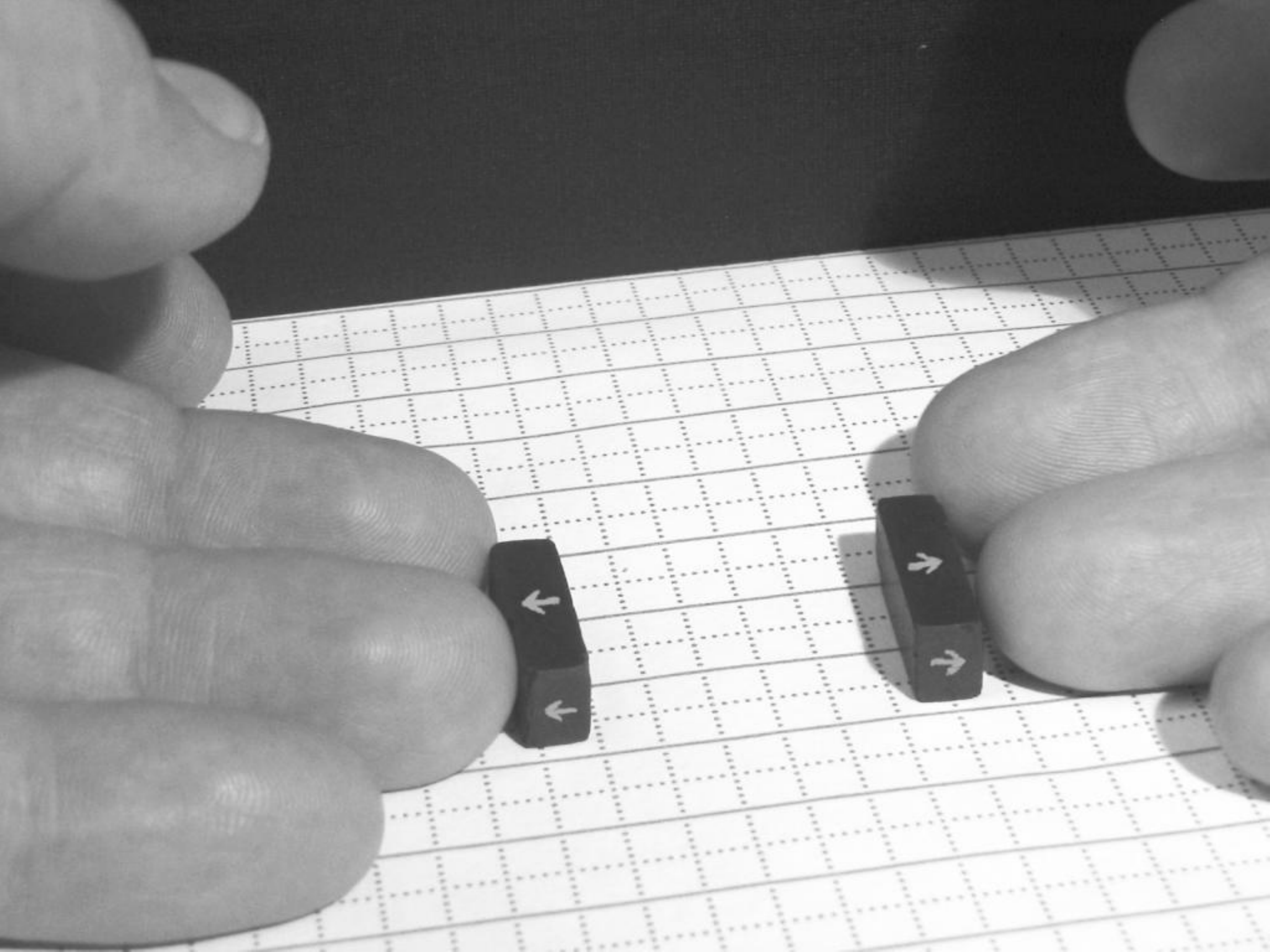}

(a)

\includegraphics[height=7.0cm]{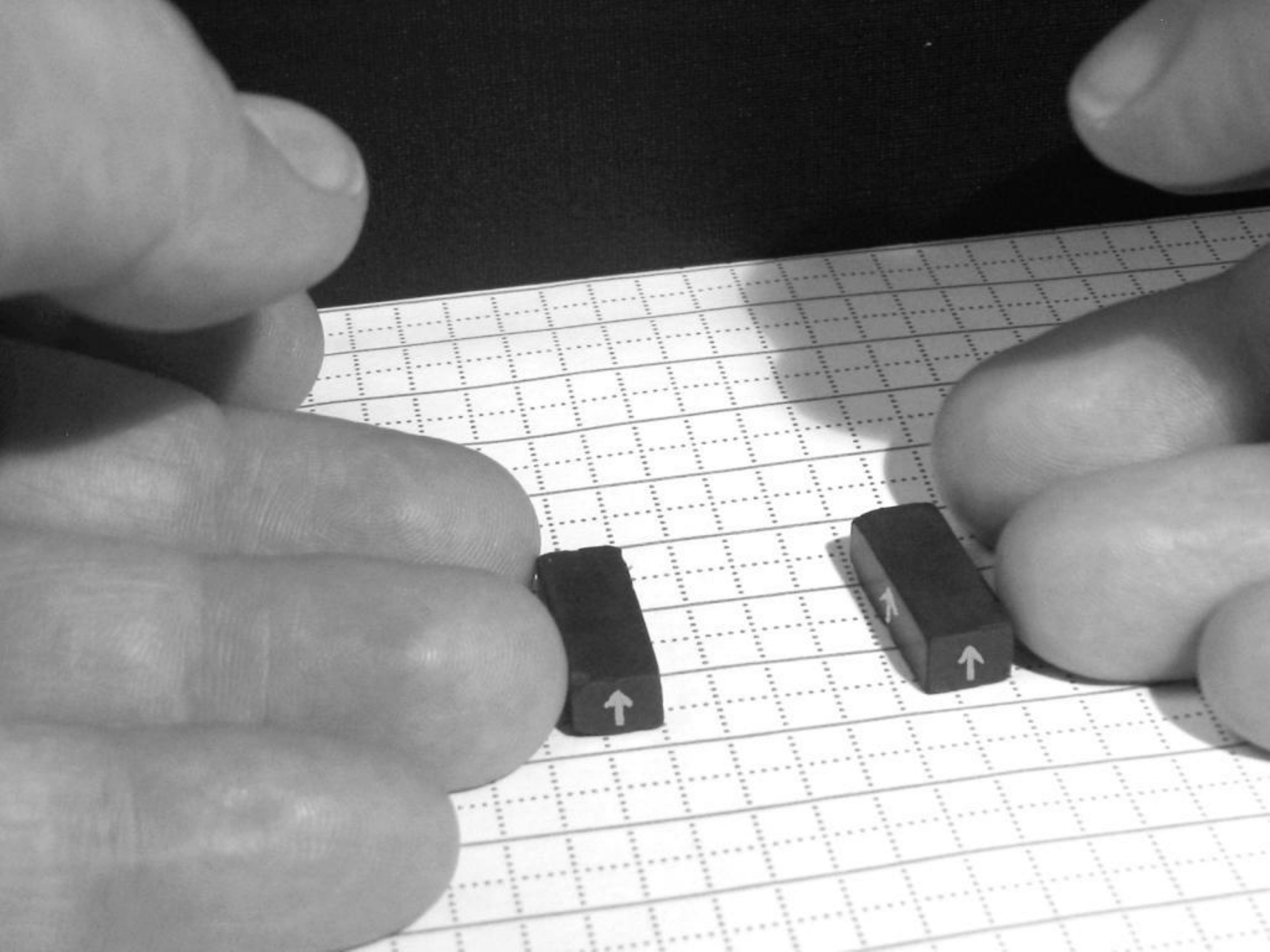}

(b)
}
 \caption{
 The magnets repelling each other are kept in position by fingers until the friction
 outweighs the repulsion (the dipole directions shown by 
 arrows): (a)  the parallel components have                                    %  parallel repulsion
 opposite direction, the stronger repulsion is felt against the friction
 to the distance of about five marks on the paper; 
 (b) the 
 perpendicular or upright components have the same direction,
  the weaker repulsion is felt against the friction
 to only about four marks on the paper.                                           %  perpendicular repulsion
}
\label{lab.fig.strong-rep}
\end{figure}

From the potential formula follows that the forces go as inverse fourth power 
of distance if the geometry is fixed. If the terms for the two configurations
differ by factor two, they will be equally strong at
distances related by fourth root of 2, i.e. 1.1892, which makes the difference about
20 per cent (18.92 per cent). This can actually be experienced, when we use the notepad paper,
as shown on the photos of figure \ref{lab.fig.strong-rep}. 
The distance at which our magnets "stop repelling",
i.e. when the friction takes over, is about 2 cm, thus the 20 per cent are 
easily seen. It should be however remarked, that the dipole formula itself
might not be fully valid due to the physical extension of the magnets
as compared to the distances involved.
It should be explained to the audience that it is used here
only for the purpose of the demonstration, while precise measurements and
analysis should be used in real studies of magnetic interactions.
%
%
%
%%%%%%%%%%%%%%%%%%%%%%%%%%%%%%%%%%%%%%....Figure 3
%
\begin{figure}[htb]
\center{
\includegraphics[height=5.5cm]{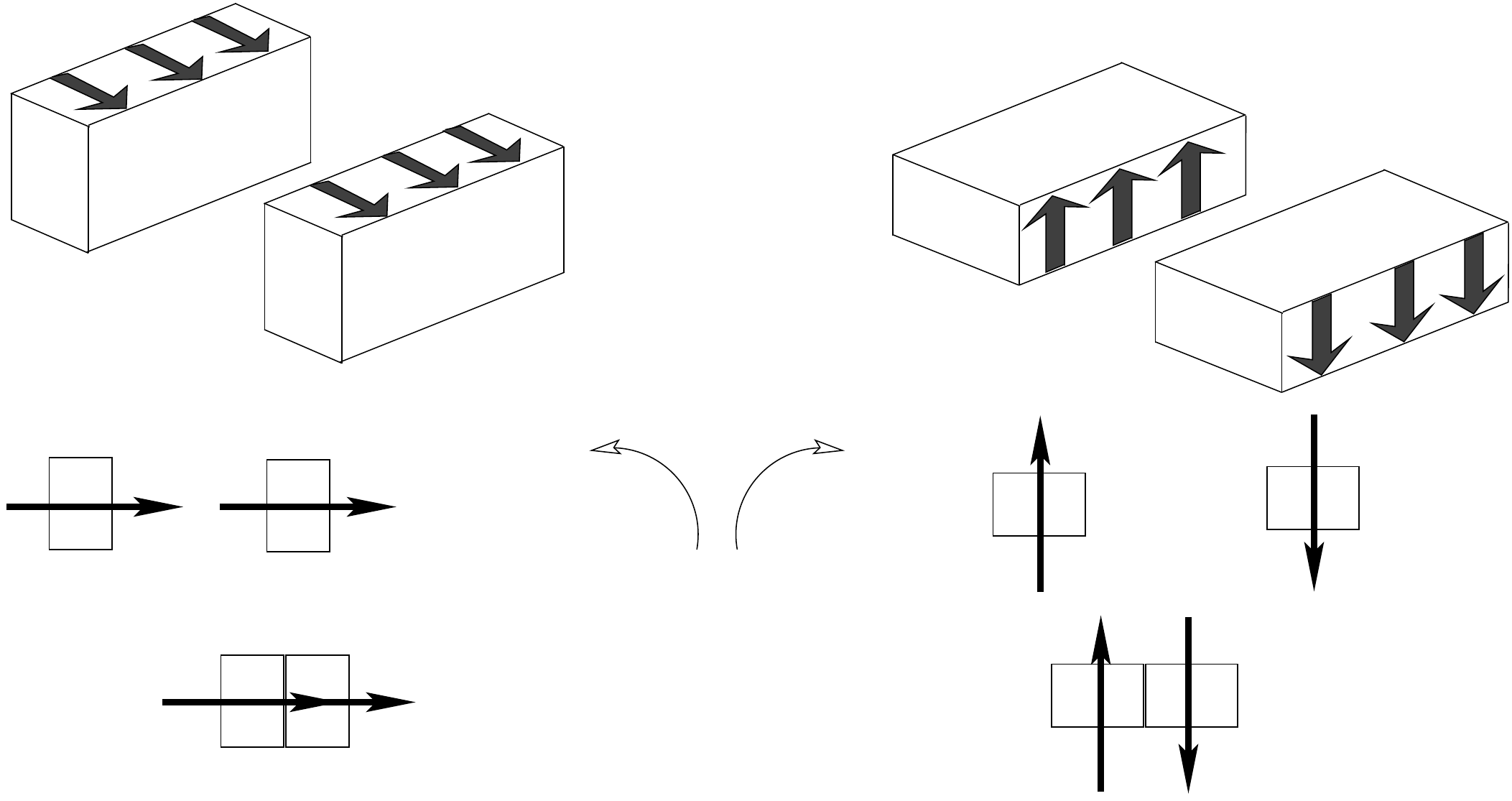}
}
 \caption{
 The magnets attracting each other. The dipol directions shown by 
 arrows.
}
 %%%
\label{lab.fig.attract}
\end{figure}
%
%
%%%%%%%%%%%%%%%%%%%%%%%%%%%%%%%%%%%%%%....Figure 4

%
%%%%%%%%%%%%%%%%%%%%%%%%%%%%%%%%%%%%%%%%
%
%%%%%%%%%%%%%%%%%%%%%%%%%%%%%%%%%%%%%%%%%%%%%%%%%%%%%%%%%%%%%
%
%
%
%
     \section{The magic angle \label{sect_magic}}  
%
%
%
%%%%%%%%%%%%%%%%%%%%%%%%%%%%%%%%%%%%%%%%%%%%%%%%%%%%%%%%%%%%%
%
%
The magic angle is used usually for the configuration where the two dipoles 
do not interact, i.e. it becomes zero.
It can happen in the case that the two dipoles are
parallel, one is placed in the origin,
both point in "the z-direction"  while their connecting 
%
%
%
%%%%%%%%%%%%%%%%%%%%%%%%%%%%%%%%%%%%%%....Figure 5
%
\begin{figure}[htb]
\center{
\includegraphics[height=7.0cm]{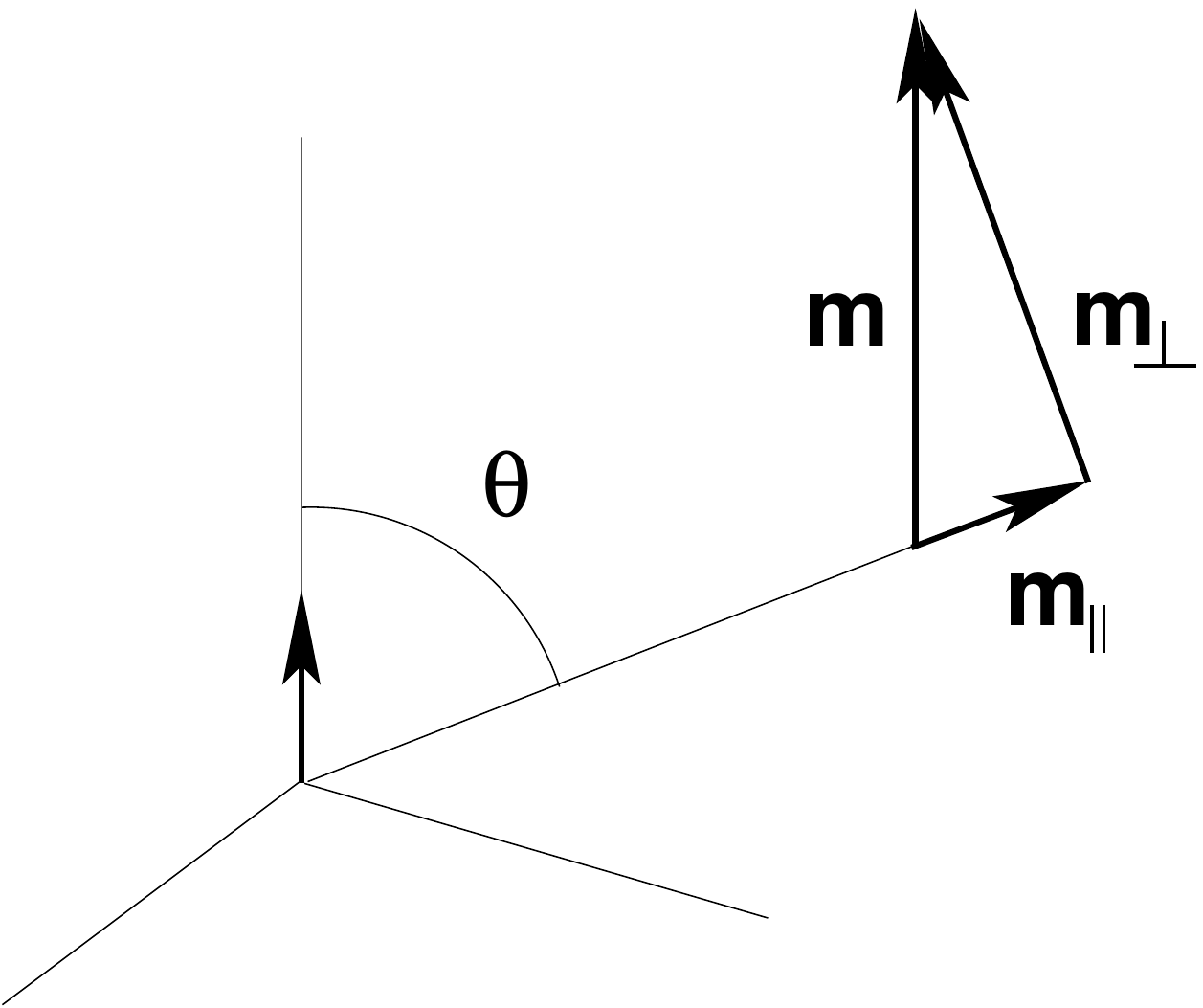}
}
 \caption{
The magic angle in the simplest configuration}
 %%%
\label{lab.fig.magicangle}
\end{figure}
%
%%%%%%%%%%%%%%%%%%%%%%%%%%%%%%%%%%%%%%%%%%%%%%%%%%%%%%%%%%%%%
vector's polar angle $\theta$ defines the parallel and perpendicular
components as 
%
%
%%%%%%%%%%%%%%%%%%%%%%%%%%%%%%%%%%%%%%%%
%
\begin{equation}
 {m}_{1 \perp}  =  {m}_{1} \sin \theta \ \ \ \ \ \
{m}_{1 \|}  =  {m}_{1} \cos \theta \ \ \ \ \ \
   \label{magic-angle_perp}
\end{equation}
%
%%%%%%%%%%%%%%%%%%%%%%%%%%%%%%%%%%%%%%%%    
%   
%
and the same for the second dipole ${m}_{2}$.
Since the dipoles are parallel, the scalar products in the paranthesis reduce to
%
%
%%%%%%%%%%%%%%%%%%%%%%%%%%%%%%%%%%%%%%%%
%
\begin{equation}
 {m}_{1 \perp}  {m}_{2 \perp} - 2 {m}_{1 \|} {m}_{2 \|}     
   \label{magic_angle_scalar_product}
\end{equation}
%
%%%%%%%%%%%%%%%%%%%%%%%%%%%%%%%%%%%%%%%%   
%   

%
%
and this gives the condition for the magic angle
%
%
%%%%%%%%%%%%%%%%%%%%%%%%%%%%%%%%%%%%%%%%
%
\begin{equation}
\sin^2 \theta  - 2  \cos^2 \theta   =  0   \ \ \ \ \ \ i.e.  \ \ \ \ \ \ 
\tan  \theta = \sqrt 2
   \label{magic_angle_condition}
\end{equation}
%
%%%%%%%%%%%%%%%%%%%%%%%%%%%%%%%%%%%%%%%%   %   
This gives the magic angle 54.74 degrees. In the literature this angle
is usually derived from the original formula where number 3 appears  
which leads simply to the condition
$ 3 \cos^2 \theta   =  1 $ giving naturally the same magic angle.
%
%
%
%%%%%%%%%%%%%%%%%%%%%%%%%%%%%%%%%%%%%%%%%%%%%%%%%%%%%%%%%%%%%
%
%
%
%
    \section{The field lines of a dipole  \label{sect_field_lines}} 
%
%
%
%%%%%%%%%%%%%%%%%%%%%%%%%%%%%%%%%%%%%%%%%%%%%%%%%%%%%%%%%%%%%
%
The formula 
for the strength of magnetic field  of the dipole $\nvec{m}$ placed at the origin
is 
%
%%%%%%%%%%%%%%%%%%%%%%%%%%%%%%%%%%%%%%%%
%
\begin{equation}
    \nvec{B}(\nvec{r})= 
    \frac{C}{ r^3}  
    \left( 
      3  \left(\nvec{m} \cdot   \frac{ \nvec{r} }{r}  \right)   \frac{ \nvec{r} }{r} 
      -
      \nvec{m} 
    \right)
    =
     \frac{C}{ r^3}  
    \left( 
      3  \left(\nvec{m} \cdot   \hat{ \nvec{r} }  \right )   \hat{ \nvec{r} } 
      -
      \nvec{m} 
    \right)   
   \label{strength_field}
\end{equation}
%
%%%%%%%%%%%%%%%%%%%%%%%%%%%%%%%%%%%%%%%%   
%   

and it invites to similar rewriting, as
%
%%%%%%%%%%%%%%%%%%%%%%%%%%%%%%%%%%%%%%%%
%
\begin{equation}
    \nvec{B}(\nvec{r})= 
    \frac{C}{  r^3}  
    \left( 
      2   \nvec{m}_{\|}
           -
           \nvec{m}_{\perp}
    \right)
   \label{strength_field_rewrite}
\end{equation}
%
%%%%%%%%%%%%%%%%%%%%%%%%%%%%%%%%%%%%%%%%    
%   
This is again much simpler, and leads to the following suggestions for the
visualizations of the field.
%
%
%%%%%%%%%%%%%%%%%%%%%%%%%%%%%%%%%%%%%%....Figure 1
%
\begin{figure}[h]
\center{
\includegraphics[height=8.5cm]{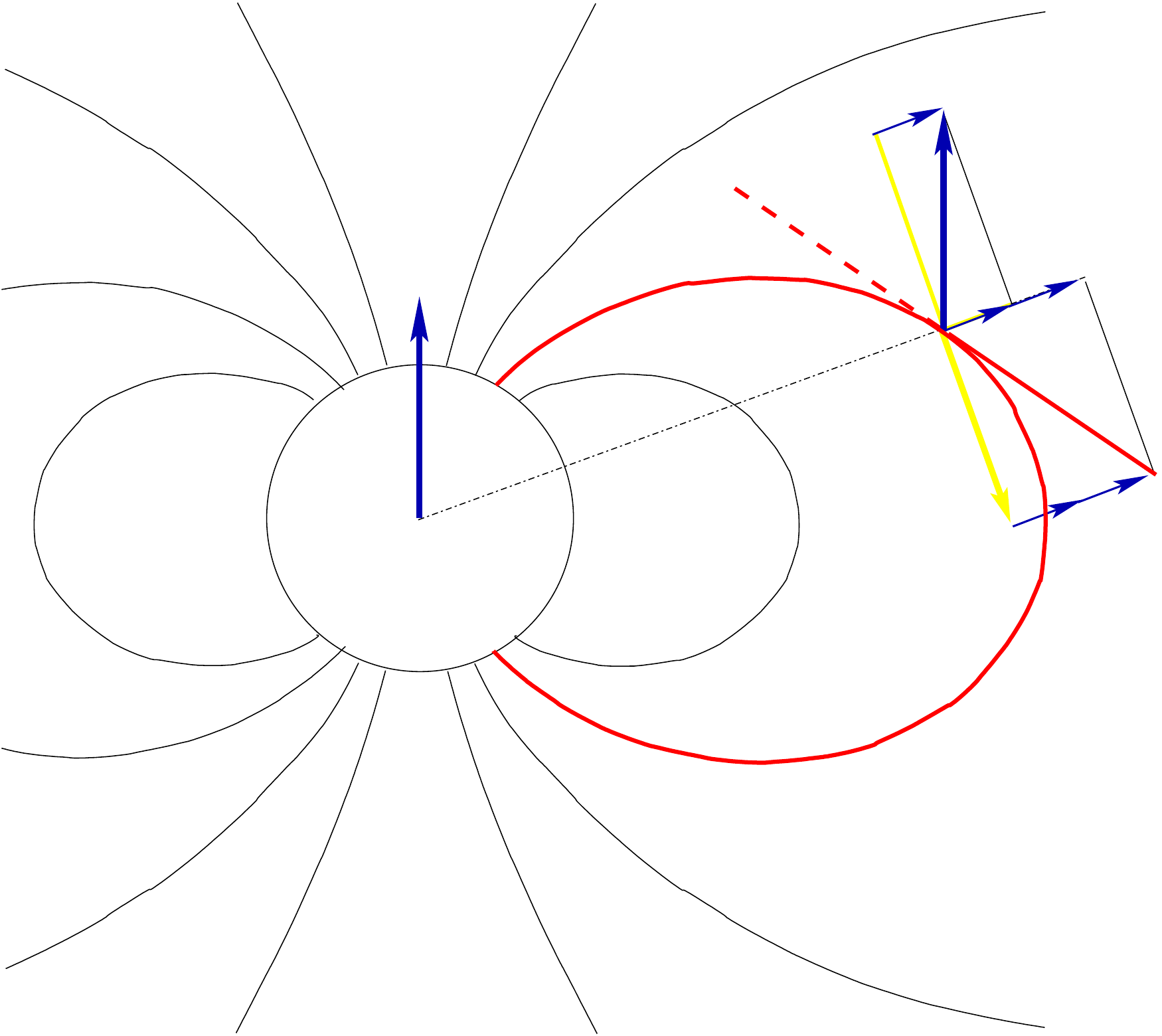}
}
 \caption{How to construct the field lines for the field of a 
 an electric or magnetic dipole. The unit vector of in the dipole direction
 is decomposed into components parallel and perpendicular to the position
 vector. Then twice the parallel and once negative of the perpendicular
 components are added to give the direction of the field, i.e. the 
 tangential line to the field line in the given position.}
 %%%
\label{field_lines}
\end{figure}
%
%%%%%%%%%%%%%%%%%%%%%%%%%%%%%%%%%%%%%%%
%
This is the prescription:
\begin{itemize}
\item
Draw the dipole vector. Make the line to a point in plane
\item
Transfer the dipole vector and decompose it in components
along and perpendicular to the connecting line
\item
Add together the twice the projection along the position vector
and add the reverse (negative)
of the perpendicular part. This gives the tangential line.
\item
Move along this direction - and start the next point
\end{itemize}
This makes fast constructins in e.g. matlab possible.
(Note: if somebody had the same idea of rewriting as presented here, 
a natural place to look for it would probably be in the source files
of simple graphics for field lines.)

The magnetic field of the dipole is also discussed at the physics
part of Eric Weisstein's science encyclopedia at Wolfram Research pages (ref. \cite{weisstein}). 
Here we have found the only place on the web 
which does not contain the factor 3, but factor 2 as
in our present discussed formula (but the entry is very brief with no discussion).
The two formulae there read
%
%%%%%%%%%%%%%%%%%%%%%%%%%%%%%%%%%%%%%%%%
%
\begin{equation}
    \nvec{B}(\nvec{r})    
    =
     \frac{3  \left(\nvec{m} \cdot \nvec{r} \right)  \nvec{r}  
      -
      \nvec{m} r^2 }
      {r^5 }
=
      \frac{m}{r^3}  
       \left(
       2 \cos \theta \hat{\nvec{r}} + 
       \sin \theta \hat{\mathbf{\theta}} 
       \right)
   \label{weisstein_eqs}
\end{equation}
%
%%%%%%%%%%%%%%%%%%%%%%%%%%%%%%%%%%%%%%%%    
%   
Here $\hat{\nvec{r}}$  and $\hat{\theta}$ are unit vectors along
the position vector and perpendicular to it, respectively. Due 
to their sign convention the minus sign essential in our discussion
is unfortunately  turned into a plus sign and the 
entry is too brief to include any interpretation. 
Note also that the units used are not specified. Fortunately, 
when compiling the final version of this paper we were
given a reference to a published discussion in ref. \cite{am_journ_disc}
 which includes correct units and a short discussion, but still without
 any interpretation along the lines of this paper.

%%%%%%%%%%%%%%%%%%%%%%%%%%%%%%%%%%%%%%%%%%%%%

\section{Taylor expansion derivation of the 
electric dipole-dipole interaction \label{sect_Taylor_expansion}} 

\begin{figure}[htb]
\center{
\includegraphics[height=4.5cm]{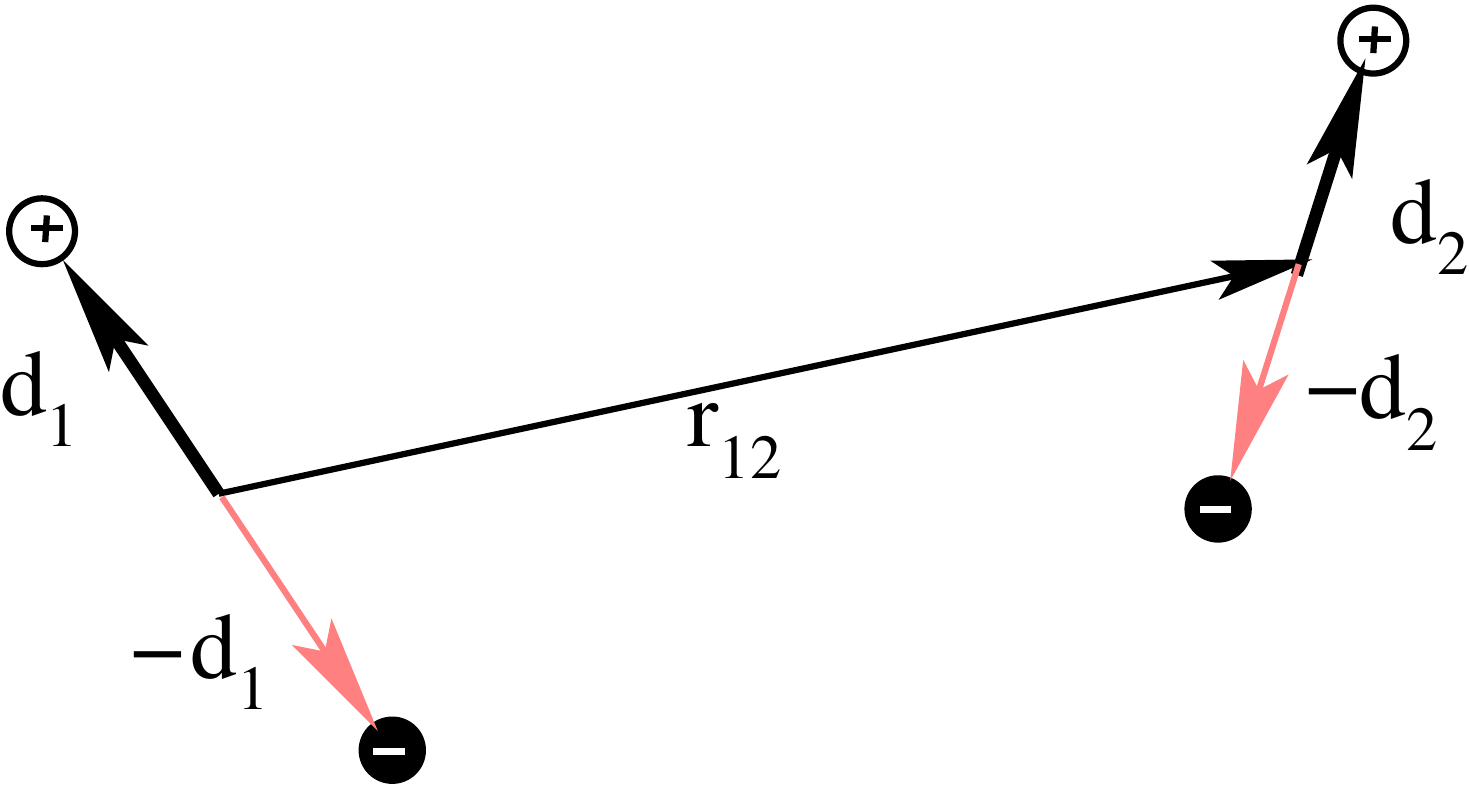}
}
 \caption{Deriving dipole-dipole potential using Taylor formula. Vectors shown by
 arrows.
}
 %%%
\label{lab.fig.dipole-taylor}
\end{figure}

The purpose of this section is to demonstrate how a very
straightforward treatment of the electric dipole-dipole interaction
from the model of two pairs of charges shown in figure \ref{lab.fig.dipole-taylor} 
leads directly to the standard dipole-dipole formula of equation \ref{standard_dipole_dipole}.
As is well known, the dipole formula is obtained in a limiting process, where
the displacements are let to tend to zero while the the product
of charge and displacement, the dipole moment, is kept constant.

The Coulomb interaction between the two pairs of charges
can be written as (now we leave the units out alltogether)
%
%%%%%%%%%%%%%%%%%%%%%%%%%%%%%%%%%%%%%%%%
%
\begin{equation}
 - \frac{q^2}      {| \nvec{r}_{12} + \nvec{d}_2 - \nvec{d}_1 |}   
 - \frac{q^2}      {| \nvec{r}_{12} - \nvec{d}_2 + \nvec{d}_1 |} 
 + \frac{q^2}      {| \nvec{r}_{12} + \nvec{d}_2 + \nvec{d}_1 |}  
 + \frac{q^2}      {| \nvec{r}_{12} - \nvec{d}_2 - \nvec{d}_1 |}  \ \ \ \ \
   \label{dipole_charges}
\end{equation}
%
%%%%%%%%%%%%%%%%%%%%%%%%%%%%%%%%%%%%%%%%   
%   
%
%%%%%%%%%%%%%%%%%%%%%%%%%%%%%%%%%%%%%%....Figure 
%
%
%%%%%%%%%%%%%%%%%%%%%%%%%%%%%%%%%%%%%%%\]
To obtain the formula for dipole-dipole interaction we need to consider
a case where  ${r}_{12}\gg d_1$ and  ${r}_{12}\gg d_1$.
Up to the second term in Taylor expansion for 2-variables:
\[ f (x + \Delta x, y + \Delta y) = f (x, y) + \left( f_x (x, y) \Delta x +
   f_y (x, y) \Delta y \right) + \]
\[ \left. \ldots + \frac{1}{2!}  \left[ f_{ {xx}} (x, y) (\Delta x)^2 + 2
   f_{ {xy}} (x, y) \Delta x \Delta y + f_{ {yy}} (x, y) (\Delta y)^2 
   \right) \right] \]
In vector notation
\[ f ( \nvec{r} + \nvec{d}) = f ( \nvec{r}) + ( \nvec{d} \cdot \nabla) ( \nvec{r})
   + \frac{1}{2!}  \nvec{d} \cdot \left[ ( \nvec{d} \cdot \nabla) (\nabla f (r))
   \right] \]
The first order terms disappear due to the symmetry of the terms, thus we only
need to consider the second order terms.

The $f (x)$ is
\[ \frac{1}{r} = \frac{1}{\sqrt{x^2 + y^2}} \]
thus we wish to obtain the vector formula for $\nvec{d} = (\Delta x, \Delta y)$
from the
very simple non-vector notation
\begin{eqnarray}
& \ & \frac{\partial^2}{\partial x_i \partial x_j} \left( \frac{1}{\sqrt{x^2 +
   y^2}} \right) = \frac{\partial^{}}{\partial x_i} \left[
   \frac{\partial^{}}{\partial x_j} \left( \frac{1}{\sqrt{x^2 + y^2}} \right)
   \right] = \frac{\partial^{}}{\partial x_i} \left[ \frac{- x_j}{\left(
   \sqrt{x^2 + y^2} \right)^3}  \right] \nonumber \\  
%   
   %\[ \frac{\partial^2}{\partial x_i \partial x_j} \left( \frac{1}{\sqrt{x^2 +
   % y^2}} \right) 
& = &     \left[ \frac{- 1}{\left( \sqrt{x^2 + y^2} \right)^3} 
   \frac{\partial^{} x_j}{\partial x_i} - x_j  \frac{\partial^{}}{\partial
   x_i} \left( \frac{1}{\left( \sqrt{x^2 + y^2} \right)^3} \right) \right] 
\nonumber \\ 
& = &   
   %   \[ \frac{\partial^2}{\partial x_i \partial x_j} \left( \frac{1}{\sqrt{x^2 +
    %    y^2}} \right) = 
   \left[ \frac{- 1}{\left( \sqrt{x^2 + y^2} \right)^3}
   \delta_{ {ij}} + 3 x_j  \left( \frac{x_i}{\left( \sqrt{x^2 + y^2}
   \right)^5} \right) \right] 
                                \label{dipole_expansion} 
\end{eqnarray}   
From this formula it is then not difficult to show, comparing with the above
vector form that Taylor expansion of
\[ 
 - \frac{q^2}      {| \nvec{r}_{12} + \nvec{d}_2 - \nvec{d}_1 |}   
 - \frac{q^2}      {| \nvec{r}_{12} - \nvec{d}_2 + \nvec{d}_1 |} 
 + \frac{q^2}      {| \nvec{r}_{12} + \nvec{d}_2 + \nvec{d}_1 |}  
 + \frac{q^2}      {| \nvec{r}_{12} - \nvec{d}_2 - \nvec{d}_1 |}
\]
where (with $\nvec{n_1}$ and $\nvec{n_2}$ the direction
unit vectors of the charge displacements in each dipole)
\[ \nvec{d}_1 = \frac{d}{2}  \nvec{n}_1 
\ \ \ \ \ \ \  
   \nvec{d}_2 = \frac{d}{2}  \nvec{n}_2 . \]
With the definitions of the dipole moments (note that the product $qd$ gives the
size of the dipole moment, which is kept constant in the limiting process, meaning that
while $d$ approaches zero, the charges $q$ must grow correspondingly)
\[ \nvec{m}_1 = (q d) \nvec{n}_1 \ \ \ \ \ \ \  
 \nvec{m}_2
   = (d q) \nvec{n}_2 \]
the above derivation in equations \ref{dipole_expansion} leads directly to 
form of the starting formula   \ref{standard_dipole_dipole}.
\[  \frac{1}{r_{12}^{\ \ 3}}  \left[ \nvec{m}_1
   \cdot \nvec{m}_2 - \frac{3 \left( \nvec{m}_1 \cdot \nvec{r}_{12}
   \right) \left( \nvec{m}_2 \cdot \nvec{r}_{12} \right)}{r_{12}^{\ \ 2}}
   \right] \]
The above  derivation 
possibly explains why the usual formula is a straightforward choice,
 and there seems no need to transform the expression. 
 This might have contributed to a general acceptance of the standard formula
 as the only reasonable choice. Also, the standard formula does not
 need any additional definition of the perpendicular components and
 appears as a most general coordinate system independent expression.
Hopefully, this paper has shown that there indeed are some advantages in 
transformations and rearrangements of the standard formula for dipole-dipole interaction
as well as the related vector fields. 
%
%
%%%%%%%%%%%%%%%%%%%%%%%%%%%%%%%%%%%%%%%%%%%%%%%%%%%%%%%%%%%%%
%
%
%
%
     \section{Conclusion \label{sect_conclude}}  
%
%
%
%%%%%%%%%%%%%%%%%%%%%%%%%%%%%%%%%%%%%%%%%%%%%%%%%%%%%%%%%%%%%
%
%
%
The dipole-dipole interaction formula has been shown to contain 
a possibility to be transformed to a much more intuitive form. We have
shown the applications which are suitable for teaching and illustrations.
It appears however that also in research applications the 
insight provided by this simple transformation of the standard formula 
might contribute to more intuitive presentations and discussions.
The transformation itself is really very elementary and it is 
thus rather surprizing that it has not been discussed earlier. 
In fact, while preparing the final version of this paper, we have 
by chance found 
a new textbook of Quantum Chemistry \cite{piela_quant_chem}
showing a  version similar to ours here, but in a component form. 
This form appears to the author to be unsafely coordinate system dependent 
and is thus immediately replaced by "waterproof" form of which is the
standard of equation \ref{standard_dipole_dipole}.

%%%%%%%%%%%%%%%%%%%%%%%%%%%%%%%%%%%%%%%%%%%%%%%%%%%%%%%%%%%%%%%%%%%
%
%
%
%
\section*{Acknowledgments}
We would like to thank Prof Lars Egil Helseth  at University of Bergen for
very helpful and enlightening discussion and suggestions.
%
%
%
%
%%%%%%%%%%%%%%%%%%%%%%%%%%%%%%%%%%%%%%%%%%%%%%%%%%%%%%%%%%%%% 
%
%
%
%%%%%%%%%%%%%%%%%%%%%%%%%%%%%%%%%%%%%%%%%%%%%%%%%%%%%%%%%%%%%%%%%%%%%%%%%%%%%%%%%%%%
\section*{References}
%%%%%%%%%%%%%%%%%%%%%%%%%%%%%%%%%%%%%%%%%%%%%%%%%%%%%%%%%%%%%%%%%%%%%%%%%%%%%%%%%%%%%

\end{document}